\documentclass[9pt]{article}
\usepackage{spconf,amsmath,graphicx,hyperref,caption, amsfonts, multirow}

\title{Fast-ULCNet: a Fast and Ultra Low Complexity Network for Single-Channel Speech Enhancement}
\name{Nicolás Arrieta Larraza, Niels de Koeijer}
\address{Bang \& Olufsen, Allé 1
7600 Struer, Denmark}
\begin{document}
%
\maketitle
\begin{abstract}
Single-channel speech enhancement algorithms are often used in resource-constrained embedded devices, where low latency and low complexity designs gain more importance. In recent years, researchers have proposed a wide variety of novel solutions to this problem. In particular, a recent deep learning model named ULCNet is among the state-of-the-art approaches in this domain. This paper proposes an adaptation of ULCNet, by replacing its GRU layers with FastGRNNs, to reduce both computational latency and complexity. Furthermore, this paper shows empirical evidence on the performance decay of FastGRNNs in long audio signals during inference due to internal state drifting, and proposes a novel approach based on a trainable complementary filter to mitigate it. The resulting model, Fast-ULCNet, performs on par with the state-of-the-art original ULCNet architecture on a speech enhancement task, while reducing its model size by more than half and decreasing its latency by 34\% on average.
\end{abstract}
\section{Introduction}
\label{sec:intro}

Single-channel speech enhancement is a key component of speech recognition, voice processing, and assistive hearing systems. Often, these technologies have real-time latency constraints and are deployed on resource-constrained embedded devices. Therefore, the use of very low complexity and low latency algorithms is needed to reduce the memory footprint and processing time, respectively. While traditional signal processing meets these requirements, deep learning has been shown to deliver superior audio quality \cite{C1}. In recent years, multiple approaches have been developed that propose novel architectures and methods to address these latency and memory constraints \cite{C2,C3,C4}. A state-of-the-art model in this regard is an architecture named ULCNet \cite{C5}. In this approach, a small model with a novel channel-wise feature reorientation block and power-law compression technique is proposed. As a result, ULCNet exhibits 3 to 4 times less computational cost and memory usage than prior state-of-the-art approaches while achieving comparable or superior noise suppression performance \cite{C5}. Gated Recurrent Units (GRUs) \cite{C6} were chosen for the Recurrent Neural Network (RNN) layers. These have been a popular choice among other low-complexity architectures \cite{C2,C3} due to their relatively low number of trainable parameters and high performance \cite{C7}.

This study proposes an extension of ULCNet, named Fast-ULCNet, which replaces the RNN units in ULCNet from GRUs to FastGRNNs. FastGRNN \cite{C8} is a popular Gated-RNN design, which achieves state-of-the-art accuracies while having 2 to 4 times fewer parameters and performing faster inferences than other RNNs. Although the authors of FastGRNN claim that the performance is invariant to the length of the input, our research shows empirical evidence that FastGRNN performance decays over time during inference of long audio signals. To solve this, we additionally propose Comfi-FastGRNN (complementary filter FastGRNN), an adaptation of FastGRNN to address performance decay over time of long sequences during the forward pass. This is achieved by incorporating a trainable complementary filter system that handles the RNN state drift, which we found to correlate with the drop in performance over time.

To the best of our knowledge, this is the first study that uses FastGRNN layers for speech enhancement and sheds light over the performance decay of FastGRNN over long input sequences. Additionally, the proposal of Comfi-FastGRNN represents the first implementation of a trainable complementary filter to deal with RNN state drift. Our results show that Fast-ULCNet achieves similar performance to the state-of-the-art architecture ULCNet, while reducing the number of parameters by more than a half and decreasing its computational latency by 34\%, on average.

The implementations of Fast-ULCNet and Comfi-FastGRNN are publicly available as open source on GitHub \footnote{\url{https://github.com/narrietal/Fast-ULCNet}}, along with an online demo \footnote{\url{https://narrietal.github.io/Fast-ULCNet/}}.

\begin{figure}[htb]

\begin{minipage}[b]{1.0\linewidth}
  \centering
  \centerline{\includegraphics[width=8.5cm]{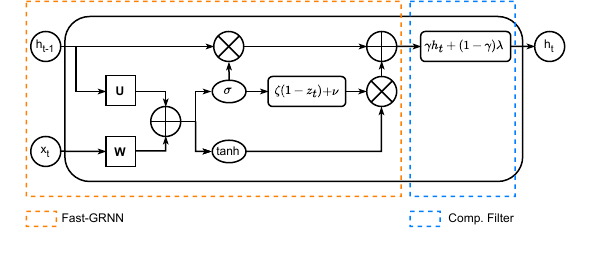}}
\end{minipage}
\caption{Block diagram of Comfi-FastGRNN, comprising the original FastGRNN architecture extended with a trainable complementary filter.}
\label{fig:comfi-fastgrnn-diagram}
\vspace{-7pt}
\end{figure}
\section{Fast-ULCNet}
\label{sec:fast-ulcnet}

\subsection{FastGRNN and Comfi-FastGRNN}
\label{ssec:comfi-fastgrnn}

FastGRNN was originally proposed as a lightweight and computationally efficient RNN architecture that delivers performance comparable to more sophisticated variants such as GRUs \cite{C8}. Its efficiency is achieved through the introduction of a weighted residual connection, implemented as a gating mechanism that reuses the same weight matrices for both the hidden-state update and the gating operation. This design not only substantially reduces the parameter count but also promotes well-conditioned gradients, thereby stabilizing the training process and alleviating the exploding and vanishing gradient problems that commonly affect conventional RNNs.

Fig.~\ref{fig:comfi-fastgrnn-diagram} illustrates the FastGRNN architecture, and (1)--(3) express its state update equations, where $\sigma$ is a non-linear activation function, $W$ and $U$ are weight matrices, and $b$ is a bias vector. Through the addition of two scalar trainable parameters $0\leq\zeta,\nu\leq1\in\mathbb{R}$, it controls the influence of the current input and previous hidden state.
 \begin{align}
z_t &= \sigma(W x_t + U h_{t-1} + b_z), \\
\tilde{h}_t &= \tanh(W x_t + U h_{t-1} + b_h), \\
h_t &= \left(\zeta (1 - z_t) + \nu \right) \odot \tilde{h}_t + z_t \odot h_{t-1}
\end{align}


 
FastGRNN achieves provably stable training, independent of the sequence length. However, this guarantees stability only during training and not in the forward pass at inference. To test the latter, the authors evaluated FastGRNN on datasets of different domains with varying sequence lengths, with the longest audio clip being 1.63 s \cite{C8}.

\begin{figure}[htb]
\begin{minipage}[b]{1.0\linewidth}
  \centering
  \centerline{\includegraphics[width=8.5cm]{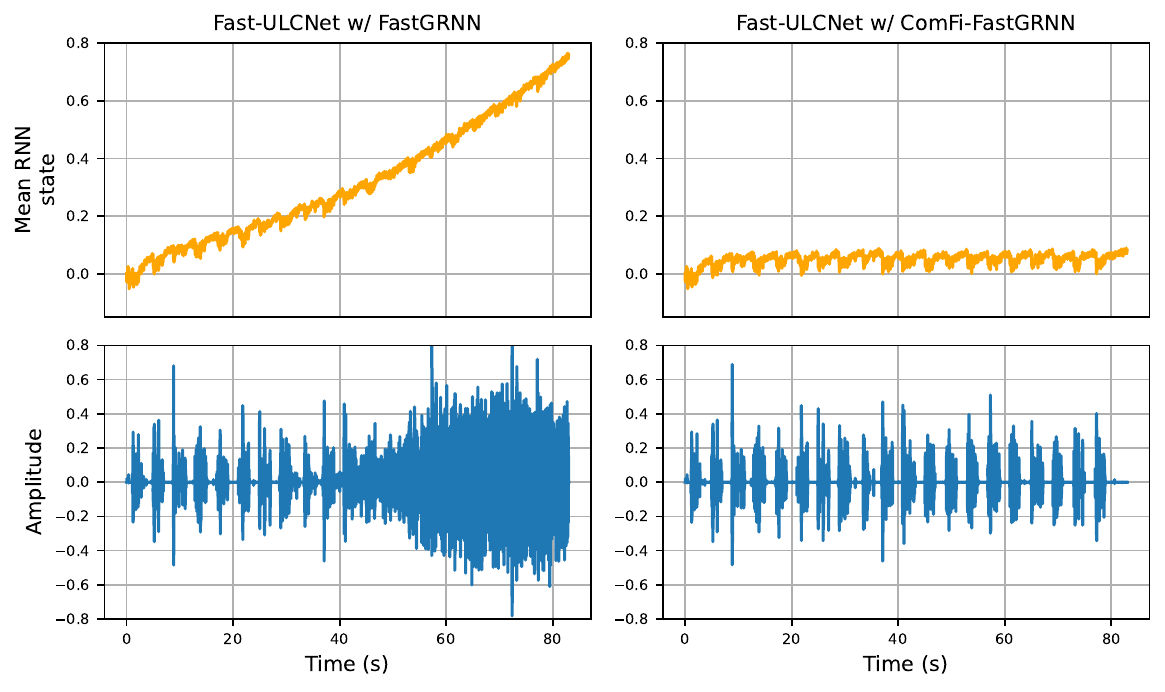}}
\end{minipage}
\caption{Fast-ULCNet inference shows drifting on the mean RNN state ${h}_t$ (top) and performance decay on the processed signal (bottom) over time with FastGRNN (left column), whereas Comfi-FastGRNN (right column) maintains stable mean RNN state $h_{t\mathrm{comfi}}$ and consistent performance.}
\label{fig:drift_vs_signal}
\vspace{-10pt}
\end{figure}

We observed that, when applying FastGRNN to longer audio signals (more than 60 s) for speech enhancement, performance degraded over time. Such a behaviour correlates with a drift in the internal RNN state, as evidenced by the increasing average hidden state magnitude over time during inference, shown in Fig.~\ref{fig:drift_vs_signal}. The drift can be traced to (3), where the internal state lacks a contraction guarantee and the coefficients do not satisfy a sum-to-one constraint, enabling state accumulation over extended inference horizons.

\vspace{-8pt}
\begin{align}
h_{t\mathrm{comfi}} &= \gamma h_t + (1 - \gamma) \lambda
\end{align}

Motivated by complementary filters used in accelerometer–gyroscope systems to reduce orientation drift, we propose Comfi-FastGRNN, an extension of FastGRNN that uses a trainable complementary filtering method to mitigate state drift. Equation (4) extends the FastGRNN state update in (3) by incorporating two trainable parameters, $\lambda,\gamma\in\mathbb{R}$. The parameter $\lambda$ acts as a scalar modulation factor to compensate for state drift, while $\gamma$ controls the relative contributions of the hidden state and the drift correction term. This solution preserves the original design intent of FastGRNN while mitigating drift through a parameter-efficient approach.


\subsection{Model architecture}
\label{ssec:model}

\begin{figure}[htb]
\begin{minipage}[b]{1.0\linewidth}
  \centering
  \centerline{\includegraphics[width=7cm]{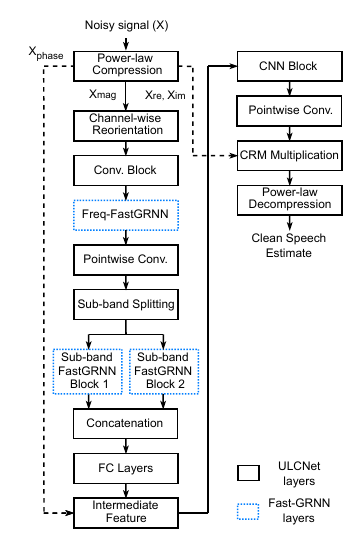}}
\end{minipage}
\caption{Architecture of Fast-ULCNet. Black boxes represent components from the original ULCNet architecture, while dotted light-blue boxes highlight the FastGRNN-based modifications introduced in this work, with or without the complementary filter. Subscripts of $X$ (phase, mag, re, and im) indicate the phase, magnitude, real, and imaginary parts of the input features, respectively.}
\label{fig:Fast-ULCNet-diagram}
\end{figure}

The proposed deep neural network architecture, illustrated in Fig.~\ref{fig:Fast-ULCNet-diagram}, is based on the ULCNet design \cite{C5}, with the GRU layers replaced by FastGRNN-based layers, implemented either as FastGRNN units or as the proposed Comfi-FastGRNN units.

In the first stage, the input features are preprocessed using a modified power-law compression applied to both real and imaginary short-time Fourier transform (STFT) components. Then, a channel-wise feature reorientation method is applied, which reduces the dimensionality of the input features for efficiency. The rest of the stage comprises a series of depthwise separable convolutional layers serving as feature extractors, followed by a bidirectional FastGRNN-based layer operating along the frequency axis to expand the receptive field. Additionally, subband-level temporal FastGRNN-based units are employed to enhance spectral modelling. A real-valued magnitude mask is subsequently predicted through a stack of two fully connected (FC) layers.

The second stage focuses on phase refinement. A convolutional neural network (CNN) is applied to intermediate representations derived from the estimated magnitude mask and the noisy phase. The final complex mask is obtained via complex ratio masking (CRM) \cite{C10}, which is used to reconstruct the enhanced complex spectrogram.

\section{Experiments}
\label{sec:experiments}

\subsection{Implementation details}
\label{ssec:implementation_details}

\subsubsection{Architecture implementation}
\label{sssec:architecture_implementation}

The ULCNet architecture was replicated in TensorFlow, adhering to the design specifications outlined by the original authors. For the channelwise feature reorientation, we apply an overlapping rectangular uniform window with a frequency resolution of 1.5 kHz and an overlap factor of 0.33. The Conv Block consists of four depthwise-separable convolutional layers with a kernel size of 1$\times$3, performing convolution operations solely along the frequency axis. These layers use 32, 64, 96, and 128 filters, respectively. Except for the first convolutional layer, downsampling is applied via max-pooling with a factor of 2 in the remaining 3 layers. The Freq-FastGRNN layer contains 64 units and is followed by a pointwise convolution with 64 filters. Subsequently, 2 subband temporal FastGRNN blocks are employed, each comprising 2 FastGRNN layers with 128 units. These are followed by 2 FC layers, each with 257 neurons. The second stage CNN consists of two 2D convolutional layers with 32 filters and a kernel size of 1$\times$3, followed by a pointwise convolutional layer with 2 output channels, which restores the desired output shape.


For the gating component $\sigma$ in our FastGRNN and Comfi-FastGRNN implementations, we adopted the sigmoid non-linearity, consistent with the original FastGRNN paper. The trainable scalar parameters of the complementary filter, $\gamma$ and $\lambda$, were initialized to 0.999 and 0.0, respectively.

\subsubsection{Loss function}
\label{sssec:loss}
We adopted a version of the Mean Absolute Error loss, which has been shown to perform favorably for speech enhancement in the time-frequency domain \cite{C11}. Specifically, our loss function comprises two components: the $L_1$ norm of the difference between the magnitude spectra of the predicted and clean speech, and the $L_1$ norm of the difference between their complex spectrogram values.

\begin{align}
\mathcal{L} = \frac{1}{T F} \sum_{t=1}^{T} \sum_{f=1}^{F} \left( \left| |S| - |\hat{S}| \right| + \left| S - \hat{S} \right| \right)
\end{align}

Equation (5) defines the loss function, where $S$ and $\hat{S}$ denote the clean and predicted spectrogram values, respectively, and $T$ and $F$ represent the total number of time frames and frequency bins. For brevity, the explicit dependence of  $S(t, f)$ and $\hat{S}(t, f)$ on time and frequency indices has been omitted.

\subsubsection{Dataset}
\label{sssec:dataset}

\begin{table*}[ht]
\centering
\caption{Objective metric results on the original 10-second DNS Challenge 2020 synthetic non-reverberant test set and a synthetically extended 90-second version. Evaluated metrics include DNSMOS (SIGMOS, BAKMOS, OVRLMOS), PESQ, and SI-SDR.}
\label{tab:model_performance}
\begin{tabular}{llccccc}
\hline
\textbf{Test signal length} & \textbf{Model} & \textbf{OVRLMOS} & \textbf{SIGMOS} & \textbf{BAKMOS} & \textbf{PESQ} & \textbf{SI-SDR} \\
\hline
\multirow{3}{*}{10 seconds} & ULCNet & \textbf{3.10} & \textbf{3.39} & 3.96 & \textbf{2.62} & \textbf{16.24} \\ 
                       & Fast-ULCNet (ours) & 3.09 & \textbf{3.39} & 3.95 & 2.51 & 15.99 \\ 
                       & Fast-ULCNet\textsubscript{comfi} (ours) & 3.09 & \textbf{3.39} & \textbf{3.97} & 2.50 & 16.01 \\ \hline
                       
\multirow{3}{*}{90 seconds} & ULCNet & 3.09 & \textbf{3.39} & 3.95 & \textbf{2.66} & \textbf{16.89} \\ 
                       & Fast-ULCNet (ours) & 2.93 & \textbf{3.39} & 3.62 & 2.24 & 13.58 \\ 
                       & Fast-ULCNet\textsubscript{comfi} (ours) & \textbf{3.10} & \textbf{3.39} & \textbf{3.99} & 2.51 & 16.48 \\ \hline
\end{tabular}
\end{table*}

For the experiments, we utilized the widely adopted Interspeech 2020 Deep Noise Suppression (DNS) Challenge dataset \cite{C12}. A total of 1000 hours of 10-second noisy speech mixtures were synthesized at a sampling rate of 16 kHz, with signal-to-noise ratio values randomly drawn from a uniform distribution ranging from \textminus10 dB to 30 dB. The dataset was then partitioned into training and validation subsets using an 85/15 split. For testing, we employed the same test set as the original ULCNet paper, which is the synthetic, non-reverberant test set provided as part of the DNS challenge.

\subsubsection{Training setup}
\label{sssec:training_setup}

Experiments were conducted using fixed hyperparameters, with training batches of 32 samples of 10 seconds each, and 4000 training steps and 1000 validation steps per epoch. The STFT used a 32-ms window, 16-ms hop size, and 512-point FFT. Optimization employed Adam with an initial learning rate of  $1 \times 10^{-3}$, gradient clipping at 3.0, and a scheduler reducing the learning rate by half after 3 epochs without validation loss improvement. Early stopping halted training if the validation loss failed to decrease for five consecutive epochs, and the model with the lowest validation loss was selected for testing.


\begin{table}[ht]
\captionsetup{skip=10pt}
\centering
\caption{Number of parameters, MACs and mean RTF measurement on the Raspberry Pi 3 B+ (Pi3) and Arm Cortex-A53 (ARM).}
\label{tab:model_performance_embedded}
\resizebox{\columnwidth}{!}{%
\begin{tabular}{lcccc}
\hline
\textbf{Model} & \textbf{Params (M)} & \textbf{MACs (M)} & \textbf{RTF\textsubscript{Pi3}} & \textbf{RTF\textsubscript{ARM}} \\
\hline
ULCNet & 0.685 & 2.057 & 0.976 & 0.927 \\
Fast-ULCNet & \textbf{0.338} & \textbf{1.691} & \textbf{0.657} & \textbf{0.604} \\
\hline
\end{tabular}
}
\end{table}
\vspace{-8pt}
\subsection{Results}
\label{ssec:results}

\subsubsection{Objective evaluation}
\label{sssec:objective_eval}

Recognizing the importance of evaluating model performance on longer audio sequences, we assess speech enhancement using two versions of the same test set. The first is the original set, containing 10-second audio samples. The second is an extended version, created by concatenating each sample with itself nine times, resulting in 90-second sequences.

Objective quality is assessed by predicting objective quality metrics. We used DNSMOS \cite{C13}, which includes the sub-metrics for speech quality (SIGMOS), background noise quality (BAKMOS), and overall quality (OVRLMOS). In addition, we report PESQ \cite{C14} and scale-invariant signal-to-distortion ratio (SI-SDR) \cite{C15} scores to provide a comprehensive evaluation.

As shown in Table~\ref{tab:model_performance}, when evaluated on the original 10-second test set, the performance of Fast-ULCNet and Fast-ULCNet\textsubscript{comfi} is comparable to that of the original ULCNet, with only minimal differences between the two Fast-ULCNet variants. This is most evident in the DNSMOS metrics, while PESQ and SI-SDR results slightly favor ULCNet.

When evaluated on the extended 90-second test set, Fast-ULCNet shows a noticeable performance drop relative to the other models, attributable to the long-term degradation effects identified in this study. In contrast, Fast-ULCNet\textsubscript{comfi} effectively mitigates this issue, achieving results on par with the original ULCNet. DNSMOS scores slightly favor this configuration, yielding modest gains in sub-metrics such as OVRLMOS and BAKMOS. However, PESQ and SI-SDR still marginally favor the original ULCNet.

\subsubsection{Computational complexity}
\label{sssec:RTF-analysis}

To evaluate the computational complexity of the models, we consider the total number of parameters, the number of multiply-accumulate operations (MACs), and the computational latency on resource-constrained platforms, measured as the mean real-time factor (RTF) over 10,000 iterations using a single thread. The embedded platforms used for this evaluation are the Arm Cortex-A53 and the Raspberry Pi 3 B+.

Table~\ref{tab:model_performance_embedded} compares the computational complexity of the ULCNet and Fast-ULCNet models. The variant with the complementary filter yields identical results and is therefore omitted for clarity. Fast-ULCNet reduces the parameter count to less than half that of ULCNet and decreases the number of MACs by 0.366 million. This reduction in complexity translates into significantly lower processing latency, with RTF improvements of approximately 33\% on the Raspberry Pi 3 B+ and 35\% on the Arm Cortex-A53.

\section{Conclusion}
\label{sec:conclusion}
In this work, we propose Fast-ULCNet, a fast and ultra-lightweight single-channel speech enhancement model. Building upon the low-complexity, state-of-the-art ULCNet architecture, we propose replacing its GRU layers with FastGRNN units to further reduce computational overhead. Additionally, we identify and empirically demonstrate a performance degradation in FastGRNN over extended time steps due to RNN state drift. To address this, we introduce Comfi-FastGRNN, an enhanced variant that incorporates a trainable complementary filter. Experimental results indicate that Fast-ULCNet achieves comparable performance to the original ULCNet, while being approximately 34\% faster and requiring less than half the parameter count. Future work may explore integrating the Comfi-FastGRNN layer into diverse architectures to assess transferability of its benefits, along with conducting perceptual evaluations across models.
\vfill
\pagebreak




\bibliographystyle{IEEEbib}
\bibliography{strings,refs}

@article{C1,
  title={Sixty years of frequency-domain monaural speech enhancement: From traditional to deep learning methods},
  author={Zheng, Chengshi and Zhang, Huiyong and Liu, Wenzhe and Luo, Xiaoxue and Li, Andong and Li, Xiaodong and Moore, Brian C.J.},
  journal={Trends in Hearing},
  volume={27},
  pages={23312165231209913},
  year={2023},
  publisher={SAGE Publications Sage CA: Los Angeles, CA}
}

@inproceedings{C2,
  title={{DeepFilterNet: A low complexity speech enhancement framework for full-band audio based on deep filtering}},
  author={Schroter, Hendrik and Escalante, Alberto N. and Rosenkranz, Tobias and Maier, Andreas},
  booktitle={ICASSP 2022 IEEE International Conference on Acoustics, Speech and Signal Processing (ICASSP)},
  pages={7407--7411},
  year={2022},
  organization={IEEE}
}

@inproceedings{C3,
  title={Real-time denoising and dereverberation wtih tiny recurrent u-net},
  author={Choi, Hyeong-Seok and Park, Sungjin and Lee, Jie Hwan and Heo, Hoon and Jeon, Dongsuk and Lee, Kyogu},
  booktitle={ICASSP 2021 IEEE International Conference on Acoustics, Speech and Signal Processing (ICASSP)},
  pages={5789--5793},
  year={2021},
  organization={IEEE}
}

@article{C4,
  title={{Low complexity speech enhancement network based on frame-level Swin transformer}},
  author={Jiang, Weiqi and Sun, Chengli and Chen, Feilong and Leng, Yan and Guo, Qiaosheng and Sun, Jiayi and Peng, Jiankun},
  journal={Electronics},
  volume={12},
  number={6},
  pages={1330},
  year={2023},
  publisher={MDPI}
}

@inproceedings{C5,
  title={Ultra low complexity deep learning based noise suppression},
  author={Shetu, Shrishti Saha and Chakrabarty, Soumitro and Thiergart, Oliver and Mabande, Edwin},
  booktitle={ICASSP 2024 IEEE International Conference on Acoustics, Speech and Signal Processing (ICASSP)},
  pages={466--470},
  year={2024},
  organization={IEEE}
}

@article{C6,
  title={On the properties of neural machine translation: Encoder-decoder approaches},
  author={Cho, Kyunghyun and Van Merri{\"e}nboer, Bart and Bahdanau, Dzmitry and Bengio, Yoshua},
  journal={arXiv preprint arXiv:1409.1259},
  year={2014}
}

@article{C7,
  title={Empirical evaluation of gated recurrent neural networks on sequence modeling},
  author={Chung, Junyoung and Gulcehre, Caglar and Cho, KyungHyun and Bengio, Yoshua},
  journal={arXiv preprint arXiv:1412.3555},
  year={2014}
}

@article{C8,
  title={{FastGRNN: A fast, accurate, stable and tiny kilobyte sized gated recurrent neural network}},
  author={Kusupati, Aditya and Singh, Manish and Bhatia, Kush and Kumar, Ashish and Jain, Prateek and Varma, Manik},
  journal={Advances in neural information processing systems},
  volume={31},
  year={2018}
}

@article{C10,
  title={Complex ratio masking for monaural speech separation},
  author={Williamson, Donald S. and Wang, Yuxuan and Wang, DeLiang},
  journal={IEEE/ACM transactions on audio, speech, and language processing},
  volume={24},
  number={3},
  pages={483--492},
  year={2015},
  publisher={IEEE}
}

@inproceedings{C11,
  title={A consolidated view of loss functions for supervised deep learning-based speech enhancement},
  author={Braun, Sebastian and Tashev, Ivan},
  booktitle={2021 44th International Conference on Telecommunications and Signal Processing (TSP)},
  pages={72--76},
  year={2021},
  organization={IEEE}
}

@article{C12,
  title={The interspeech 2020 deep noise suppression challenge: Datasets, subjective testing framework, and challenge results},
  author={Reddy, Chandan K.A. and Gopal, Vishak and Cutler, Ross and Beyrami, Ebrahim and Cheng, Roger and Dubey, Harishchandra and Matusevych, Sergiy and Aichner, Robert and Aazami, Ashkan and Braun, Sebastian and others},
  journal={arXiv preprint arXiv:2005.13981},
  year={2020}
}

@inproceedings{C13,
  title={{DNSMOS P. 835: A non-intrusive perceptual objective speech quality metric to evaluate noise suppressors}},
  author={Reddy, Chandan K.A. and Gopal, Vishak and Cutler, Ross},
  booktitle={ICASSP 2022 IEEE international conference on acoustics, speech and signal processing (ICASSP)},
  pages={886--890},
  year={2022},
  organization={IEEE}
}

@techreport{C14,
  author      = {{ITU-T}},
  title       = {{Recommendation P.862: Perceptual Evaluation of Speech Quality (PESQ): An Objective Method for End-to-End Speech Quality Assessment of Narrow-band Telephone Networks and Speech Codecs}},
  institution = {{International Telecommunication Union}},
  year        = {2001},
  type        = {Standard},
  number      = {P.862}
}

@inproceedings{C15,
  title={{SDR--half-baked or well done?}},
  author={Le Roux, Jonathan and Wisdom, Scott and Erdogan, Hakan and Hershey, John R},
  booktitle={ICASSP 2019 IEEE International Conference on Acoustics, Speech and Signal Processing (ICASSP)},
  pages={626--630},
  year={2019},
  organization={IEEE}
}

\end{document}